\documentclass[aps,preprint]{revtex4}
\usepackage{amssymb}
\usepackage{graphicx}
\usepackage{amsmath}

\begin{document}

\title{Bessel functions of integer order in terms of hyperbolic functions}

\author{V. B\^{a}rsan}
\affiliation{IFIN, Bucharest-Magurele, Romania.}
\author{S. Cojocaru}
\affiliation{Universit{\`a} degli Studi di Salerno, Italy \\
Institute of Applied Physics, Chi\c{s}in\u{a}u, Moldova}

\begin{abstract}
Using Jacobi's identity, a simple formula expressing Bessel functions of
integer order as simple combinations of powers and hyperbolic functions,
plus higher order corrections, is obtained.
\end{abstract}

\maketitle

In this article we shall propose a simple formula expressing the
modified Bessel functions of integer order, $I_{n},$ in terms of
powers and hyperbolic functions of the same argument. It can be
easily adapted for the Bessel functions $J_{n}.$

The starting point is the generalization of the Jacobi identity (\cite{W}
p.22) used in the calculation of lattice sums \cite{1}:

\begin{equation}
\frac{1}{N}\sum_{m=0}^{N-1}\exp \left[ \frac{z}{2}\left( we^{i\frac{2\pi m}{N%
}}+\frac{1}{w}e^{-i\frac{2\pi m}{N}}\right) \right] =\sum_{k=-\infty
}^{\infty }w^{kN}\cdot I_{kN}(z).  \label{1}
\end{equation}

For $w=1$ and $N=2$, (\ref{1}) gives a well-known formula:

\begin{equation}
\cosh z=\sum_{k=-\infty }^{\infty }I_{2k}(z)=I_{0}(z)+2\left[
I_{2}(z)+I_{4}(z)+...\right] ,  \label{2}
\end{equation}
(see for instance \cite{2}, eq.9.6.39)

It is interesting to exploit (\ref{1}) at $w=1$, for larger values of $N$.
For $N=4$,

\begin{equation}
\frac{1}{2}\left( 1+\cosh z\right) =\cosh ^{2}\frac{z}{2}=I_{0}(z)+2%
\sum_{k=1}^{\infty }I_{4k}(z),  \label{3}
\end{equation}
and for $N=8:$

\begin{equation}
\frac{1}{4}(1+\cosh z+2\cosh \frac{z}{\sqrt{2}})=I_{0}(z)+2\sum_{k=1}^{%
\infty }I_{8k}(z).  \label{4}
\end{equation}

It is easy to see that, if the l.h.s. of these equations contains $p$
hyperbolic cosines, it provides an exact expression for the series of $%
I_{0}(z)$, cut off at the $z^{4p}$ term. The generalization of (\ref{4}) for
$N=4p,$ with $p$ - an arbitrary integer$,$ is indeed:

\begin{equation}
\frac{1}{2p}\left[ 1+\cosh z+2\cosh \left( z\cos \frac{\pi }{2p}\right)
+...+2\cosh \left( z\cos \left( p-1\right) \frac{\pi }{2p}\right) \right]
=I_{0}(z)+2\sum_{k=1}^{\infty }I_{4pk}(z).  \label{5}
\end{equation}

Because

\begin{equation}
I_{n}(-iz)=i^{-n}J_{n}(z).  \label{6}
\end{equation}
our result (\ref{5}) can be written as:

\begin{equation}
\frac{1}{2p}\left[ 1+\cos z+2\cos \left( z\cos \frac{\pi }{2p}\right)
+...+2\cos \left( z\cos \left( p-1\right) \frac{\pi }{2p}\right) \right] =
\label{7}
\end{equation}

\begin{equation*}
=J_{0}(z)+2\sum_{k=1}^{\infty }J_{4pk}(z).
\end{equation*}

It is easy to obtain formulae similar to (\ref{5}) for any modified Bessel
function of integer order. Let us introduce the notations:

\begin{equation}
c_{1}=\cos \frac{\pi }{2p},\qquad ...\qquad c_{p-1}=\cos \frac{p-1}{2p}\pi .
\label{8}
\end{equation}
and let us define the functions:

\begin{equation}
S_{q}\left( z\right) =\sinh z+2\left( c_{1}\right) ^{q}\sinh \left(
c_{1}z\right) +...+2\left( c_{p}\right) ^{q}\sinh \left( c_{p}z\right)
,\qquad q>0  \label{9}
\end{equation}

\begin{equation}
C_{q}\left( z\right) =\cosh z+2\left( c_{1}\right) ^{q}\cosh \left(
c_{1}z\right) +...+2\left( c_{p}\right) ^{q}\cosh \left( c_{p}z\right)
,\qquad q\geq 0  \label{10}
\end{equation}

We have:

\begin{equation}
\frac{dS_{q}\left( z\right) }{dz}=C_{q+1}\left( z\right) ;\qquad \frac{%
dC_{q}\left( z\right) }{dz}=S_{q+1}\left( z\right)  \label{11}
\end{equation}

Using recurrently the formula (\cite{Gradst} 8.486.4):

\begin{equation}
\frac{d}{dz}I_{n}\left( z\right) -\frac{n}{z}I_{n}\left( z\right)
=I_{n+1}\left( z\right)  \label{12}
\end{equation}
and the notation:

\begin{equation}
T_{n}\left( z\right) =\left( \frac{1}{z}\frac{d}{dz}\right) ^{n}C_{0}\left(
z\right) ,  \label{13}
\end{equation}
we get:

\begin{equation}
\frac{1}{2p}z^{n}T_{n}=I_{n}\left( z\right) +2z^{n}\left( \frac{1}{z}\frac{d%
}{dz}\right) ^{n}\sum_{k=1}^{\infty }I_{4pk}(z).  \label{14}
\end{equation}
\qquad

For $n=1,2,3,4:$

\begin{equation}
T_{1}\left( z\right) =z^{-1}S_{1}\left( z\right) ,\text{ }%
T_{2}=-z^{-3}S_{1}\left( z\right) +z^{-2}C_{2}\left( z\right) \text{ ,}
\label{15}
\end{equation}

\begin{equation}
T_{3}\left( z\right) =3z^{-5}S_{1}\left( z\right) -3z^{-4}C_{2}\left(
z\right) +z^{-3}S_{3}\left( z\right) ,  \label{16}
\end{equation}

\begin{equation}
T_{4}\left( z\right) =-15z^{-7}S_{1}\left( z\right) +15z^{-6}C_{2}\left(
z\right) -6z^{-5}S_{3}\left( z\right) +z^{-4}C_{4}\left( z\right) .
\label{17}
\end{equation}

The general expressions are:

\begin{equation}
T_{2n}=z^{-2n}\left[ \alpha _{1}^{\left( 2n\right) }z^{-2n+1}S_{1}+\alpha
_{2}^{\left( 2n\right) }z^{-2n+2}C_{2}+...+\alpha _{2n}^{\left( 2n\right)
}C_{2n}\right] ,  \label{18}
\end{equation}

\begin{equation}
T_{2n+1}=z^{-2n-1}\left[ \alpha _{1}^{\left( 2n+1\right)
}z^{-2n}S_{1}+\alpha _{2}^{\left( 2n+1\right) }z^{-2n+1}C_{2}+...+\alpha
_{2n+1}^{\left( 2n+1\right) }S_{2n+1}\right] ,  \label{19}
\end{equation}

We get:

\begin{equation}
\alpha _{1}^{\left( n\right) }=-\alpha _{2}^{\left( n\right) }=\left(
-1\right) ^{n+1}\left( 2n-3\right) !!,\qquad \alpha _{n-1}^{\left( n\right)
}=-\frac{\left( n-1\right) n}{2},\qquad \alpha _{n}^{\left( n\right) }=1.
\label{20}
\end{equation}%
and the following recurrence relations for the coefficients $\alpha
_{q}^{\left( p\right) }:$

\begin{equation}
\alpha _{n-p}^{\left( n\right) }=\alpha _{n-p-1}^{\left( n-1\right) }-\left(
n+p-2\right) \alpha _{n-p}^{\left( n-1\right) },\qquad 2\leqslant p\leqslant
n-3.  \label{21}
\end{equation}

Other general expressions of the coefficients are:

\begin{equation}
\alpha _{3}^{\left( n\right) }=\left( -1\right) ^{n+1}\left( n-2\right)
\left( 2n-5\right) !!  \label{22}
\end{equation}

\begin{equation}
\alpha _{4}^{\left( n\right) }=\left( -1\right) ^{n}\left( n-3\right)
!\left\{ \frac{2^{n-4}}{0!}1!!+\frac{2^{n-3}}{1!}3!!+...+\frac{2^{0}}{\left(
n-4\right) !}\left( 2n-7\right) !!\right\} .  \label{23}
\end{equation}

Ignoring the series in the r.h.s. of (\ref{14}), we get approximate
expression for $I_{n}.$ Let us give here these expressions for the value $%
p=2 $ and $n=0,1,2,3:$

\begin{equation}
I_{0}^{\left( ap\right) }(z)=\frac{1}{4}\cdot \left( 1+\cosh z+2\cosh \frac{z%
}{\sqrt{2}}\right) ,  \label{24}
\end{equation}

\begin{equation}
I_{1}^{\left( ap\right) }\left( z\right) =\frac{1}{4}\cdot \left( \sinh z+%
\sqrt{2}\sinh \frac{z}{\sqrt{2}}\right) ,  \label{25}
\end{equation}

\begin{equation}
I_{2}^{\left( ap\right) }\left( z\right) =\frac{1}{4}\cdot \left( -\frac{1}{z%
}\left( \sinh z+\sqrt{2}\sinh \frac{z}{\sqrt{2}}\right) +\cosh z+\cosh \frac{%
z}{\sqrt{2}}\right) ,  \label{26}
\end{equation}
\begin{equation*}
I_{3}^{\left( ap\right) }\left( z\right) =\frac{1}{4}\left[ \frac{3}{z^{2}}%
\left( \sinh z+\sqrt{2}\sinh \frac{z}{\sqrt{2}}\right) -\frac{3}{z}\left(
\cosh z+\cosh \frac{z}{\sqrt{2}}\right) \right.
\end{equation*}

\begin{equation}
\left. -\frac{3}{z}\left( \cosh z+\cosh \frac{z}{\sqrt{2}}\right) +\left(
\sinh z+\frac{1}{\sqrt{2}}\sinh \frac{z}{\sqrt{2}}\right) \right] .
\label{27}
\end{equation}

According to the Table 1, even for this very small value of $p$, the
approximation provided by these functions, for "moderate" values of the
argument $(z\lesssim 4),$ is very good.

\bigskip

Table 1

\begin{tabular}{|c|c|c|c|c|}
\hline
$z$ & $1$ & $2$ & $3$ & $4$ \\ \hline
$\frac{I_{0}^{\left( ap\right) }-I_{0}}{I_{0}}$ & $1.6\times 10^{-7}$ & $%
2.4\times 10^{-5}$ & $3.3\times 10^{-4}$ & $1.7\times 10^{-3}$ \\ \hline
$\frac{I_{1}^{\left( ap\right) }-I_{1}}{I_{1}}$ & $2.3\times 10^{-6}$ & $%
1.4\times 10^{-4}$ & $1.2\times 10^{-3}$ & $4.4\times 10^{-3}$ \\ \hline
$\frac{I_{2}^{\left( ap\right) }-I_{2}}{I_{2}}$ & $7.1\times 10^{-5}$ & $%
10^{-3}$ & $4.5\times 10^{-3}$ & $1.2\times 10^{-2}$ \\ \hline
$\frac{I_{3}^{\left( ap\right) }-I_{3}}{I_{3}}$ & $1.8\times 10^{-3}$ & $%
7.3\times 10^{-3}$ & $1.7\times 10^{-2}$ & $3\times 10^{-2}$ \\ \hline
\end{tabular}

\bigskip

It is visible that the precision of the approximation decreases with the
order $n$ of the Bessel function $I_{n}$. We can increase it arbitrarly, by
increasing the value of $p.$

The extension of these results to Bessel functions of real argument is
trivial, using the formula (\ref{6}) and:

\begin{equation}
S_{q}\left( -iz\right) =-i\left[ \sin z+2c_{1}^{q}\sin \left( c_{1}z\right)
+...+2c_{p}^{q}\sin \left( c_{p}z\right) \right] ,  \label{28}
\end{equation}

\begin{equation}
C_{q}\left( -iz\right) =\cos z+2c_{1}^{q}\cos \left( c_{1}z\right)
+...+2c_{p}^{q}\cos \left( c_{p}z\right) .  \label{29}
\end{equation}

In conclusion, we have proposed a controlled analytic approximation for
Bessel functions of integer order. The first $4p-n$ terms of the series
representation of $I_{n}$ is generated exactly by the first $4p-n$ terms of
the elementary functions in the l.h.s. of eq.(14). So, our formulae can be
used, for instance, to find the series expansions of the powers of Bessel
functions, a subject discussed recently by Bender et al \cite{Bender}

The results presented in this paper can be applied to a large variety of
problems, mainly with cylindrical symmetry, involving Bessel functions at
"moderate" arguments. They may provide also a useful "visualization" of $%
J_{n}$ and $I_{n}$ in terms of elementary functions. The method cannot be
used for asymptotic problems.

\end{document}